\documentclass[aps,prl,floatfix,groupedaddress,,nofootinbib,notitlepage,twocolumn]{revtex4-2}
\usepackage{times,bm,bbm,bbold,amssymb,amsmath,amsfonts,dsfont,cancel,graphics,graphicx,color}
\usepackage[normalem]{ulem}
\usepackage[usenames,dvipsnames]{xcolor}
\usepackage[colorlinks=true,citecolor=blue,linkcolor=blue,urlcolor=blue]{hyperref}
\DeclareFontFamily{OT1}{pzc}{}
\DeclareFontShape{OT1}{pzc}{m}{it}
              {<-> s * [1.25] pzcmi7t}{}
\DeclareMathAlphabet{\mathpzc}{OT1}{pzc}
                                 {m}{it}

\usepackage[T1]{fontenc} 
\usepackage{newtxtext,newtxmath} 

\def \[[{{[\hskip-1mm[}}
\def \]]{{]\hskip-0.9mm]}}

\begin{document}
\title{State-Based Quantum Simulation of Imaginary-Time Evolution}
\author{ S. Alipour}
\email{s.alipoor@gmail.com}
\affiliation{Computational Physics Laboratory, Physics Unit, Faculty of Engineering and Natural Sciences, Tampere University, FI-33014 Tampere, Finland}
\author{ T. Ojanen}
\affiliation{Computational Physics Laboratory, Physics Unit, Faculty of Engineering and Natural Sciences, Tampere University, FI-33014 Tampere, Finland}

%%%%%%%%%%%%%%%%%%%%%%%%%%%%%%%%%%%%%%%%%%%%%%%%%%%%%%%%%%%%%%%%
\begin{abstract}
Imaginary time evolution is a powerful technique for computing the ground state of quantum Hamiltonians, where the convergence to ground state in asymptotic imaginary time is guaranteed. However, implementing this method on quantum computers is challenging due to its nonunitary nature. Here, we propose a fully quantum approach for simulation of imaginary time evolutions which eliminates the need for intermediate classical computation or state tomography. Our method leverages the recently introduced state-based quantum simulation technique, in which using quantum states beside quantum gates allows to simulate a broader class of evolutions beyond the natural quantum dynamics. Specifically, we demonstrate how by using a set of quantum states and by applying only controlled-$\textsc{swap}$ gates and measurements, one can simulate the nonunitary imaginary time evolution. We illustrate our results through an example. 
\end{abstract}
\maketitle
%%%%%%%%%%%%%%%%%%%%%%%%%%%%%%%%%%%%%%%%%%%%%%%%%%%%%%%%%%%%%%%%
\textit{Introduction.---}Many problems in various areas of science from machine learning to computational biology concern some sort of optimization that can be translated into finding the ground state of a Hamiltonian \cite{DFT, qchemistry, GS-based_qfeature_map, qbiology}. As such, ground-state simulation and preparation are always of paramount importance and impacts on various domains spanning physics, chemistry, computer science, and beyond. 

In numerical simulations on classical computers, imaginary-time evolution is a highly successful methods for computing ground states \cite{MonteCarlo,CITE}. By a Wick rotation which replaces real time with imaginary time in a quantum evolution, quantum mechanics is connected to statistical mechanics, which enables studying finite temperature properties of a quantum system \cite{Wick, QFT-ITE-1,QFT-ITE-2,MPS-ITE-1, MPS-ITE-2, MPS-ITE-3}. In addition, studying low-temperature properties of a system and finding its ground state become possible by finding the trajectory of the imaginary-time-evolved state \cite{MonteCarlo,PDE-ITE}. 

Quantum computers can potentially simulate this evolution with an exponential speedup compared to the classical computers. However, since imaginary-time evolution is not unitary and hence is considered as a nonphysical process, it is not possible to simulate it naturally on a quantum computer. There are a number of other quantum algorithms for finding the ground state on quantum computers. However, since this problem is QMA-complete, it is hard to find the solution even on quantum computers \cite{Terhal-QMA,Kempe-QMA,Aharonov-QMA}. The existing algorithms include phase estimation algorithm, which despite recent progresses, still has drawbacks for implementation in near-term quantum computers \cite{KitaevQPE, CEMM98, Lin:QPE2023, PhaseEstimationBased-1,PhaseEstimationBased-2,PhaseEstimationBased-3,PhaseEstimationBased-4, Aspuru}. In addition to this, finding the ground state can be done through quantum annealing, which relies on adiabatic changes of a suitable time-dependent Hamiltonian, or through variational methods such as variational eigensolvers \cite{qAnnealing,qAnnealingIsing,AQC-Farhi,QOA-Farhi,VQES, Kandala-Nature, Yuan-Theory_VQS,q.natural-gradient}. The latter is a hybrid classical-quantum algorithm where part of the computation runs on a classical computer. However, both approaches have limitations; quantum annealing can be slow and there is no guarantee that variational methods converge to the ground state. 

Due to these challenges related to existing quantum algorithms, there is an increasing interest to investigate possibilities for implementation of imaginary-time evolution on quantum computers.  In imaginary-time evolution convergence to the ground state is certain in the asymptotic limit and there is no need for limiting the speed of the evolution. 

To address the inherent nonunitarity issue of this evolution, several approaches have been proposed in recent years \cite{VQITE,Brandao:QITE, Nishi:QITE}. One approach relies on variational techniques. In this approach, a parametrized quantum state is generated by a quantum circuit composed of parametrized quantum gates, which is used as an ansatz. This ansatz is then optimized to approximate the imaginary-time-evolved state. A key challenge in all variational methods is that the parametrized state may not cover all possible states in the corresponding Hilbert space. This can lead to deviation of the optimized state from the actual solution of the problem \cite{VQITE}. 
Another proposed method is unitrization, where the nonunitary evolution of an initial state is replaced by a sequence of unitary operations. These unitary transformations are designed such that their concatenation reproduces the trajectory of the original imaginary-time evolution. However, such a unitary dynamics becomes state-dependent and the need for tomography becomes unavoidable, unless at the cost of substantial classical computation. 
This limits the effectiveness of this method \cite{Brandao:QITE, Nishi:QITE}. 

Here we propose a new technique for quantum simulation of imaginary-time evolution. For implementation of this technique, we do not use any classical computation or optimization and we do not need to tomography the state of the system in the intermediate time. To achieve this fully quantum mechanical simulation, we use state-based quantum simulation (SBQS) technique \cite{SBQS}. In contrast to the conventional simulation methods, where the Hamiltonian is decomposed in terms of local Hamiltonian operators \cite{book:Nielsen-Chuang, Preskill, Abrams-Lloyd-1, Abrams-Lloyd-2,qsim-entropy,Nori}, in the SBQS method we decompose the Hamiltonian in terms of a set of quantum density matrices. We demonstrate that by applying controlled-\textsc{swap} gates on these states and a set of qubit ancilla prepared in suitable states, in addition to the later performances of appropriate measurements, any imaginary-time Hamiltonian dynamics can be simulated by this method. 

\textit{Imaginary-time evolution.---}Imaginary-time evolution of a system with Hamiltonian $H$ is described by imaginary-time Schr\"{o}dinger equation as
\begin{align}
\partial_{\beta} \sigma(\beta)=-\{H,\sigma(\beta)\}.
\end{align}
The solution of this equation at a given imaginary time $\beta$ is given by 
\begin{align}
\label{eq2}
\sigma(\beta)=e^{-\beta H}\sigma(0) e^{-\beta H},
\end{align}
 which is an unnormalized quantum state. Normalizing this state $\widetilde{\sigma}(\beta)=\sigma(\beta)/\mathrm{Tr}[\sigma(\beta)]$, it is easy to see that 
 \begin{align}
\lim_{\beta \to \infty} \widetilde{\sigma}(\beta)=  |E_0\rangle\langle E_0|,
 \end{align}
where $|E_0\rangle$ is the ground state of $H$ and $E_0$ is the ground state energy. This relation shows that the ground state of the Hamiltonian is achievable in the asymptotic limit of $\beta$. In the following, we show how to simulate this evolution on quantum computers using the SBQS method. 
 
 \textit{SBQS of imaginary-time evolution.---}In the SBQS method, quantum states, rather than quantum gates, have the primary role in quantum simulation. The main essence of this method is decomposition of the given known quantum Hamiltonian in terms of a set of quantum states \{$\varrho_i$\} such that
\begin{align}
\label{H_decomp}
H=\textstyle{\sum_{i=1}^{\ell} h_i \varrho_i},
\end{align}
and $\ell$ is the number of terms in the decomposition. There are no restrictions on how to choose this set of states and they can be chosen simply according to available resources in the lab for quantum state preparation. Due to this, these states are called resource states. 

Given the decomposition of Eq.~\eqref{H_decomp}, we prepare a set of quantum systems in the resource states $\varrho_i$ and a set of auxiliary qubit systems. In the SBQS method, auxiliary qubits are mainly used as control systems for applying control gates on the target systems and are prepared in suitable pure states as
\begin{align}
\label{control-qubit}
 |\psi_{\delta_i}\rangle=|0\rangle - \delta_i |1\rangle,
\end{align}
which is normalized up to $O(\delta^2)$. The coefficient $\delta_i$ in auxiliary qubit states are determined based on the coefficients $h_i$ in the Hamiltonian decomposition of Eq.~\eqref{H_decomp}.  Next, inspired by the density matrix exponentiation technique \cite{Lloyd-Mohseni-DME,DME-experimental}, we apply a series of controlled-\textsc{swap} gates and measurements on the resource systems, auxiliary qubits, and a simulator system which is initially in the state $\sigma_0$ with Hilbert-space dimention $d$. 
Without loss of generality, we assume $\sigma_0$ is pure, i.e., $\sigma_0= |\sigma_0\rangle\langle \sigma_0|$. In the following we provide the detailed recipe for application of the  controlled-\textsc{swap} gates and measurements on the quantum systems involved in the simulation process with the aim of generation of the imaginary-time evolved state $ e^{-\beta H}|\sigma_0 \rangle $ on the simulator.

To simulate this dynamics we first decompose the evolution into $N$ imaginary-time steps, i.e., $e^{-\beta H}=(e^{-\frac{\beta}{N} H})^N$. 
We then use a product formula known as the Trotter-Suzuki expansion, which decomposes the evolution generated by $H$ into the concatenation of evolutions generated by each term in the Hamiltonian decomposition,
\begin{align}
\label{Trotter-Suzuki}
(&e^{-\frac{\beta}{N} H})^N =\textstyle{ (\Pi_{i=1}^{\ell}e^{-\frac{\beta h_i }{N} \varrho_i})^N +\varepsilon},
\end{align}
where $\epsilon_{\textsc{ts}}=O\big(\frac{1}{2N}\binom{\ell}{2}\max_{ i,j}\|[\beta h_i\varrho_i, \beta h_j\varrho_j]\|\big)$ is the error \cite{book:Rivas-Huelga, Trotter-Suzuki-1,Trotter-Suzuki-2,eBook:product formulae}.
Since $\|\varrho_i\| \leqslant 1$, then  $\|[h_i\varrho_i, h_j\varrho_j]\|\leqslant 2|h_i h_j| \leqslant 2| h_{\max}|^2$, hence
\begin{align}
\label{Trotter-Suzuki-error}
\varepsilon= O(\frac{\ell^2 \beta^2 | h_{\max}|^2}{N}).
\end{align}

To simulate our desired evolution we need to simulate each term in the Trotter-Suzuki equation \eqref{Trotter-Suzuki}. We define 
\begin{align}
\delta_i:= \frac{\beta h_i }{N},
\end{align}
and choose $N \gg 1$ such that $\delta_i$ and $\epsilon_{\textsc{ts}}$ are sufficiently small. To simulate the first term $e^{-\delta_1 \varrho_1}$ in Eq.~\eqref{Trotter-Suzuki} we first prepare a control qubit in the state  $|\psi_{\delta_1}\rangle$ introduced as in Eq.~\eqref{control-qubit}. We then apply a controlled-\textsc{swap} gate, $\mathpzc{U}_{\textsc{cs}_1}$, on the control qubit $\psi_{\delta_1}=|\psi_{\delta_1}\rangle \langle\psi_{\delta_1}|$ in addition to two target systems which are the resource system $\varrho_1$ and the simulator system $\sigma_0$. We then discard the resource system by tracing it out. This leads to the quantum state $\xi_1$ given by 
\begin{align}
\label{step-before-M}
\xi_1 & =\mathrm{Tr}_{\textsc{r}_1} [ \mathpzc{U}_{\textsc{cs}_1} (\psi_{\delta_1} \otimes \varrho_1 \otimes \sigma_0) \mathpzc{U}_{\textsc{cs}_1}^{\dag} ] \nonumber\\
 & =(|0 \rangle \otimes | \sigma_0 \rangle - \delta_1 |1 \rangle \otimes  \varrho_1 |\sigma_0\rangle)(\mathrm{h.c.})+ O(\delta_1^2),
\end{align}
in which the subscript $\textsc{r}_1$ indicates that $\mathrm{Tr}$ is applied on resources system $1$.
Now we can take two different strategies: 

\textit{First strategy: Performing measurement.---}To complete the simulation of the first term in Eq.~\eqref{Trotter-Suzuki}, we perform a selective measurement in the $|\pm\rangle$ basis on the control qubit and keep the state only when the measurement output is $|+\rangle$, i.e., 
\begin{align}
M_{+}\,\xi_1 M_{+}= & |+ \rangle\langle +| \otimes \big( \sigma_0 -  \delta_1 \{ \varrho_1, \sigma_0\})/2+ O(\delta_1^2),
\end{align}
where  $M_+= |+\rangle\langle+|$.
After discarding the control system we obtain the updated state $\sigma_1$ of the simulator system which is equivalent to simulation of one step of the dynamics of the first term in Eq.~\eqref{Trotter-Suzuki},
\begin{align*}
\sigma_1 =(\sigma_0 -  \delta_1 \{ \varrho_1, \sigma_0\})/2 &\approx (\mathbbmss{I} -\delta_i\varrho_i) \sigma_1 (\mathbbmss{I} -\delta_i\varrho_i)\nonumber\\
&= e^{-\delta_1\varrho_1} \sigma_0 \, e^{-\delta_1 \varrho_1}/2 + O(\delta_1^2).
\end{align*}
It is important to note that $\sigma_1$ is not normalized. Hence the simulator system after this process takes the following state 
\begin{align}
\widetilde{\sigma}_1=\sigma_1/\mathrm{Tr}[\sigma_1],
\end{align}
with probability
\begin{align}
p_1=\mathrm{Tr}[\sigma_1]= \mathrm{Tr}[e^{-\delta_1 \varrho_1}\sigma_0 \, e^{-\delta_1 \varrho_1}]/2.
\end{align}

Now we similarly simulate the evolution of the second term $e^{-\delta_2 \varrho_2}$ on $\widetilde{\sigma}_1$  using the resource state $\varrho_2$ and control qubit $|\psi_{\delta_2}\rangle$, to obtain the the second step updated simulator system $\widetilde{\sigma}_2=\sigma_2 / \mathrm{Tr}[\sigma_2]$. Here $\sigma_2= e^{-\delta_2 \varrho_2} \widetilde{\sigma}_1 e^{-\delta_2 \varrho_2}/2$ and $\widetilde{\sigma}_2$ is obtained with probability 
\begin{align}
p_2=\mathrm{Tr}[\sigma_2]=\mathrm{Tr}[e^{-\delta_2 \varrho_2} e^{-\delta_1\varrho_1} \sigma_0 \, e^{-\delta_1 \varrho_1} e^{-\delta_2 \varrho_2}]/(2^2 p_1).\nonumber
\end{align}
 %============
\begin{figure}[tp]
\includegraphics[width=0.8\linewidth]{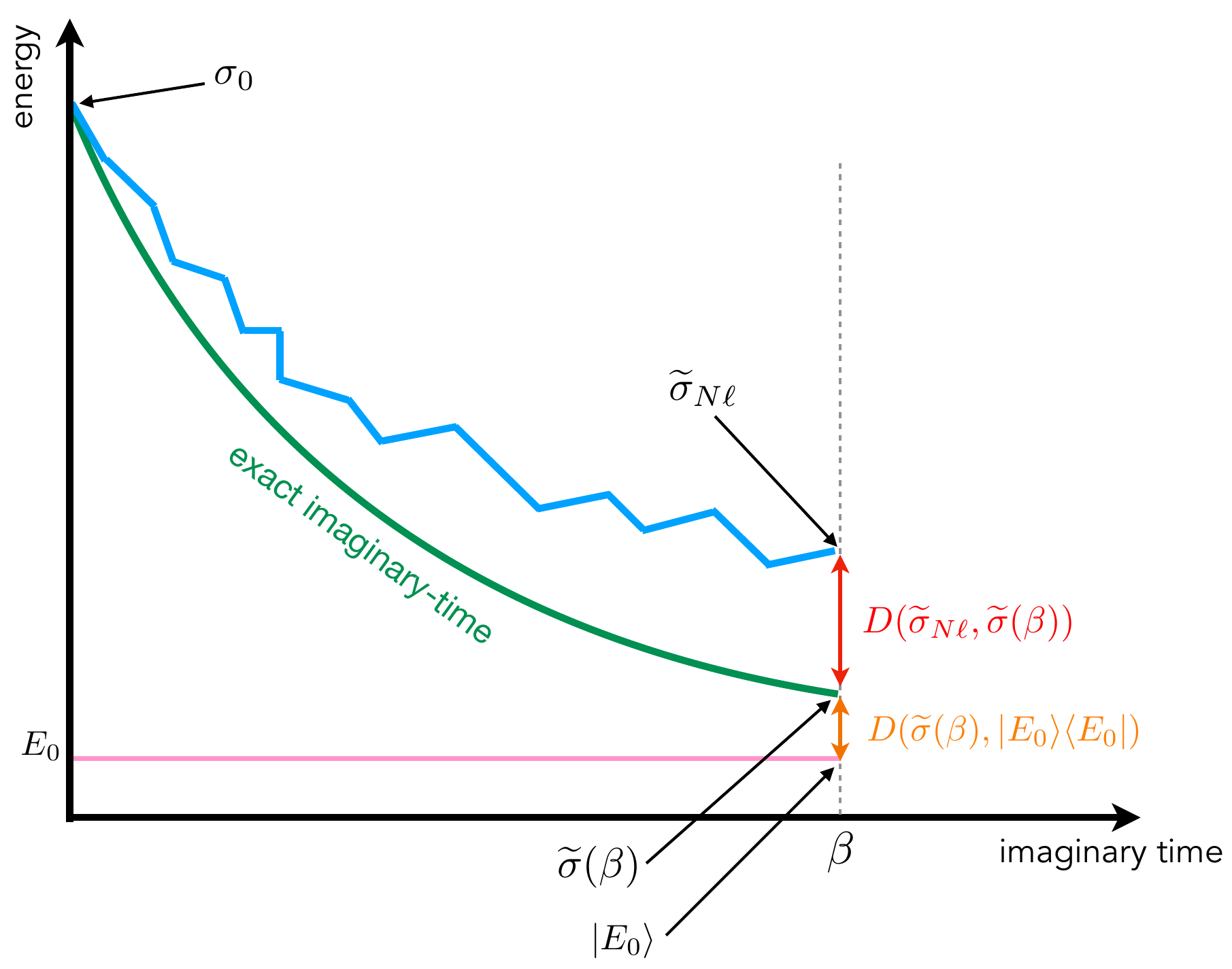}
\caption{Schematic diagram of energy vs. imaginary time errors as the distance between the state generated by the simulated dynamics (blue) and the ground state, and the distance between the exact imaginary-time evolved state (green) and the ground state in a finite imaginary time $\beta$. 
}
\label{fig:fidelity}
\end{figure}
%============
\textit{Error analysis.---}Following similar steps for all $\ell$ terms of Eq.~\eqref{Trotter-Suzuki} we can simulate the process operator $\Pi_{i=1}^{\ell}e^{-\delta_i \varrho_i}$ by approximating it with $\Pi_{i=1}^{\ell}(\mathbbmss{I} -\delta_i\varrho_i)$ with a truncation error of the order of (see SM)
 \begin{align}
O(\ell \max_i \|e^{-\delta_i \varrho_i}\|^{\ell-1} \max_i \{\delta_i^2\})= O(\ell \delta^2),
 \end{align}
where $\delta \equiv \max_i \{\delta_i\}$. The right-hand side holds because $\|e^{-\delta_i \varrho_i}\|^{\ell-1} \leqslant e^{{\| \delta_i \varrho_i \|}^{\ell-1}}$ and  $\delta^{\ell-1} \propto (1/N)^{\ell-1} \ll 1$, thus $\|e^{-\delta_i \varrho_i}\|^{\ell-1} \approx 1 $.

 By repeating the whole  process $N$ times, the final simulator state $\widetilde{\sigma}_{N\ell} \approx \widetilde{\sigma}(\beta)=e^{-\beta H} \sigma_0 e^{-\beta H} /\mathcal{N}$, with $\mathcal{N}=\mathrm{Tr}[ e^{-2\beta H} \sigma_0]$ can be achieved by 
 generating the evolution $(\Pi_{i=1}^{\ell}e^{-\delta_i \varrho_i})^N$ with an error 
 of the order of 
 \begin{align}
O(N \ell \delta^2 \max_i \|e^{-\delta_i \varrho_i}\|^{\ell(N-1)} ).
 \end{align}
 
 \textit{Remark}.---We can always assume that $H$ is a positive operator. That is, any nonpositive Hamiltonian can be shifted to a positive one so that its imaginary-time evolution leads to the same trajectory as of $H$ up to a normalization factor (details in SM).  Therefore, we can always find a decomposition such that $h_i$s are positive and hence $\delta_i>0$. Thereby $\|e^{-\delta_i \varrho_i}\| < 1$, which leads to the an upper bound for the error as
$
O(N \ell \delta^2).
$
Summing up this error for implementation of $(\Pi_{i=1}^{\ell}e^{-\delta_i \varrho_i})^N$ and $\varepsilon$, the error of the Trotter-Suzuki expansion \eqref{Trotter-Suzuki-error}, we can find the error of the SBQS simulation as 
\begin{align}
\label{err-sim-sigmabeta}
D\big(\widetilde{\sigma}_{N\ell} , \widetilde{\sigma}(\beta)\big)= O\big(2 \ell^2 \beta^2 h^2/N\big),
\end{align}
where $D$ up to a prefactor $\sqrt{2}$ is the Bures distance which is connected to fidelity  $F(\sigma_1, \sigma_2)=\left(\mathrm{Tr}\left[\sqrt{\sqrt{\sigma_1} \sigma_2 \sqrt{\sigma_1}}\right]\right)^2$ by $D^2=2(1-\sqrt{F})$, and we have defined the shorthand $h \equiv \max_i \{h_i\}$.
It can be shown that fidelity of $\widetilde{\sigma}(\beta)$ and the ground state $|E_0\rangle$ is equal to $F(\widetilde{\sigma}(\beta), |E_0\rangle\langle E_0|)=\langle E_0| \widetilde{\sigma}(\beta)| E_0\rangle$ which leads to (see SM)
\begin{align}
F \big(\widetilde{\sigma}(\beta ), |E_0\rangle\langle E_0| \big)
\geqslant  1/ \big(1+d\, e^{-2\beta \Delta} /F_0 \big),
\end{align}
where $F_0$ is fidelity of the initial simulator state and the ground state of the Hamiltonian $F_0= \langle E_0| \sigma_0 |E_0\rangle$, and $\Delta= E_1-E_0$ is the energy gap between the first excited-state and the ground state energy of $H$. Form this, it can be  obtained that the distance of the imaginary time evolved state $\widetilde{\sigma}(\beta)$ at $\beta$ with the ground state, satisfies the following inequality:
$
D^2(\widetilde{\sigma}(\beta), |E_0\rangle\langle E_0|)  
\leqslant 
1-\left(1+ e^{-2\beta \Delta} \frac{1-F_0}{F_0} \right)^{-1}.
$
As expected, if the initial simulator state is $|E_0\rangle$ the upper bound becomes in the above inequality vanishes. 
To achieve an estimate of the ground state to a desired error $\epsilon$, it is sufficient to have (see SM)
\begin{align}
\frac{2}{N} \ell^2 \beta^2 h^2+  \sqrt{ 1-\left(1+ e^{-2\beta \Delta} \frac{1-F_0}{F_0} \right)^{-1}}\leqslant \epsilon.
\end{align}
To obtain an estimation of the number of steps for simulation $N^{\ast}$ and the value of $\beta^{\ast}$ to satisfy the desired error, it is sufficient that each term on the left-hand-side is smaller than $\epsilon/2$, hence
\begin{align}
\beta &\geqslant \frac{1}{2\Delta}\log \left(\frac{(1-F_0)(4-\epsilon^2)}{F_0 \, \epsilon^2}\right) =: \beta^{\ast},\\
N & \geqslant
4 \ell^2 \beta^2 h^2 \epsilon^{-1} \geqslant
 \frac{\|H\|^2}{\epsilon \Delta^2 } \left[\log \left(\frac{(1-F_0)(4-\epsilon^2)}{F_0 \, \epsilon^2}\right)\right]^2 =: N^{\ast},\nonumber
\end{align}
where we have used $\|H\|\leqslant \ell h$.

\textit{Probability.---}After a total of $N \ell$ steps of the simulation the final simulator state $\widetilde{\sigma}_{N\ell} \approx \widetilde{\sigma}(\beta)=e^{-\beta H} \sigma_0 e^{-\beta H} /\mathcal{N}$ with $\mathcal{N}=\mathrm{Tr}[ e^{-2\beta H} \sigma_0]$ can be achieved with probability $p=\Pi_{i=1}^{N\ell} p_i$, such that
\begin{align}
p \leqslant p^{\ast}:= 2^{-N^{\ast}\ell}\, \mathrm{Tr}[e^{-\beta^{\ast} H} \sigma_0 e^{-\beta^{\ast} H}],
\end{align}
where $p^{\ast}$ is the probability of obtaining the final state within the desired error $\epsilon$ from the ground state.
We can summarize the whole error and probability analysis as 
\begin{align}
\mathrm{Prob}\big[D(\widetilde{\sigma}_{N\ell}, |E_0\rangle\langle E_0|) \leqslant \epsilon \big] \leqslant p^{\ast}.
\end{align}

\textit{Second strategy: Deferring the measurements to the end.---}In this strategy after the step in Eq.~\eqref{step-before-M} no measurement is applied. We rather apply the next controlled-\textsc{swap} gate $\mathpzc{U}_{\textsc{cs}_2}$ with a new control qubit $|\psi_{\delta_{2}}\rangle$ and the evolved simulator state as the target state, which leads to
\begin{align}
\xi_2 &= \mathrm{Tr}_{\textsc{r}_2}[\mathpzc{U}_{\textsc{cs}_2} (\psi_{\delta_{2}} \otimes  \varrho_2 \otimes \xi_1) \mathpzc{U}_{\textsc{cs}_2}^{\dag}]\nonumber\\
&= (|00 \rangle \otimes  |\sigma_0\rangle -  \delta_1 |01 \rangle \otimes  \varrho_{1} |\sigma_0 \rangle- \delta_{2} |10\rangle \otimes \varrho_{2} |\sigma_0 \rangle)(\mathrm{h.c.})\nonumber\\
&~~~+O(\delta_1^2+\delta_2^2).
\end{align}
 %============
\begin{figure*}[tp]
\includegraphics[width=0.8\linewidth]{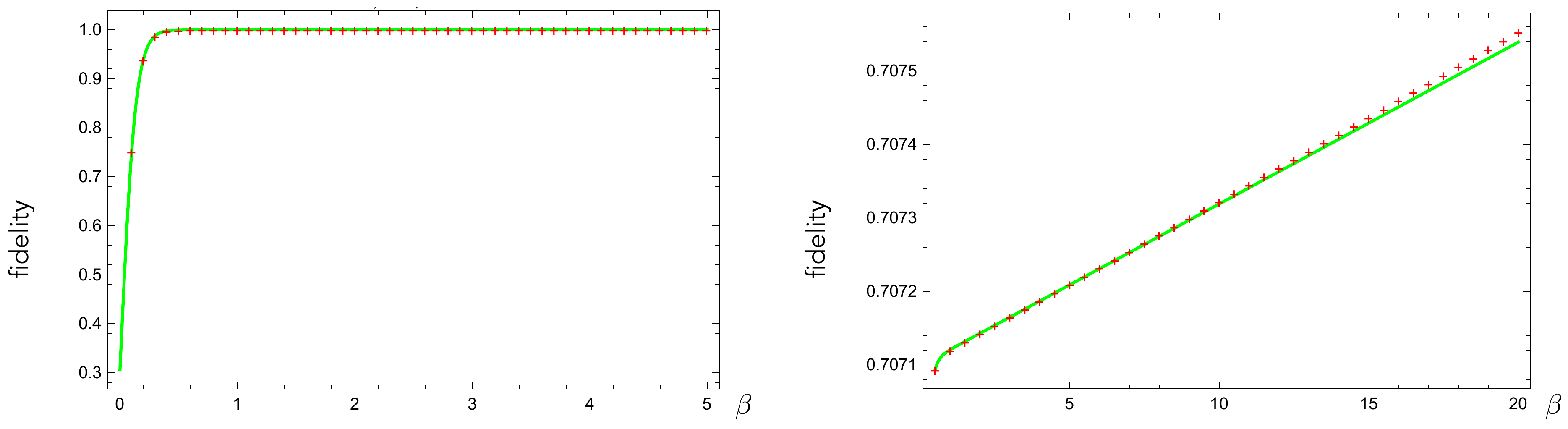}
\caption{The green solid plots: Fidelity of the successful SBQS imaginary-time evolution with the exact ground state of the Hamiltonian of Eq.~\eqref{H_Ising} vs. imaginary time $\beta$, when $J=1$ and  $|\sigma_0\rangle$ is the uniform state for (left) far from the degenerate point with $B=5$ and (right) close to the degenerate point $B=0.1$. 
The red ``+'' plots: Fidelity of the exact imaginary-time evolution with the exact ground state of the same Hamiltonian, as a benchmark, when the corresponding values of the couplings and the initial state are similar to the SBQS case. 
}
\label{fig:fidelity}
\end{figure*}
%============
Applying all $\ell$ necessary controlled-\textsc{swap} gates and tracing over all reourse systems, we can find the total state of the whole system of controlled qubits and the simulator as 
\begin{align}
\label{xi_ell}
&\xi_{\ell} \approx (|\bm{0}\rangle \otimes |\sigma_0\rangle - \textstyle{\sum_{i=1}^{\ell}} \delta_i |\bm{i} \rangle \otimes  \varrho_{i} |\sigma_0 \rangle)(\mathrm{h.c.}),
\end{align}
in which $|\bm{i}\rangle$ is a quantum state in the Hilbert space of $\ell$ control qubits with $0$ in all sites except a $1$ in $i$th cite, i.e., $ |\bm{i}\rangle= |\underbrace{0,\cdots, 0}_{i-1}, 1, \underbrace{0,\cdots, 0}_{\ell-i}\rangle$,
and $|\bm{0}\rangle$ is the state with $0$ at all sites. 
 We now apply measurements on $\xi_{\ell}$ aiming to obtain the state $\sigma_0 - \sum_{i=1}^{\ell} \delta_i \{\varrho_i , \sigma_0\} \approx e^{-\sum_i \delta_i \varrho_i} \sigma_0 e^{-\sum_i \delta_i \varrho_i}$. If we apply projector $M_+=|+ \rangle\langle +|^{\otimes \ell}$ on control qubits  in $\xi_\ell$, which is equivalent to post-selecting the results of the measurement when all control qubits are in the state $|+\rangle$, we get to the desired state with probability $\mathcal{N}/2^{\ell}$. We, however, can improve the success probability by performing a global measurement with POVMs $M= (\frac{1}{\sqrt{\ell+1}}\sum_{i=0}^{\ell} |\bm{i}\rangle)(\mathrm{h.c.})$ and $\mathbbmss{I}-M$. If we post-select only the results related to POVM $M$, the success probability becomes $\mathcal{N}/(\ell+1) $ which is an exponential improvement compared to the local measurements and $\mathcal{N}/2^{\ell+1}$. 
 
 The error of the simulation in this case is equivalent to the error of approximation $\Pi_{i=1}^{\ell}e^{ \delta_i \varrho_i} \approx \mathbbmss{I}+\sum_{i=1}^{\ell} \delta_i \varrho_i $, and is of the order of $O(\ell^2 \delta^2)$ (see SM). Repeating this process $N$ times related to $N$ imaginary-time steps in the Trotter-Suzuki expansion, the success probability becomes
 \begin{align}
 p= (\ell+1)^{-N} \, \mathrm{Tr}[e^{-\beta H} \sigma_0 e^{-\beta H}],
 \end{align}
 and error is upper bounded with Eq.~\eqref{err-sim-sigmabeta} similar to the last scenario. 
 
It should be noted that, in this $N$ times repetition, we cannot use the same trick by postponing measurements to the end and applying a global measurement. This is due to the reverse proportionality of $\delta_i$ and $N$ which does not allow neglecting terms like $\delta_i^N$ (for details see SM). 

\textit{Remark.---}It can be shown by using the median lemma that if success probability in a simulation step is larger than $1/2$, it can be improved to a desired probability by increasing the number of repetitions in that step of the simulation \cite{Wiebe-PoweringLemma}. In our case, this may entail some tradeoff between the final simulation error and its success probability. In addition, it seems possible to improve this probability by using more sophisticated measurement schemes, such as performing measurements in $\delta$-dependent basis rather than in the fixed $|\pm\rangle$ basis. 
   
\textit{Example: One dimensional Ising model in transverse field.---} 
Assume a one dimensional chain of $N$ spin-$1/2$ particles with Ising Hamiltonian in transverse field 
\begin{align}
\label{H_Ising}
\textstyle{H=-J \sum_{i=-m}^{m} X_{i} X_{i+1}- B \sum_{i=-m}^{m}   Z_{i}.}
\end{align}
in which $J$, coupling, and $B$ the amplitude of the external field are real numbers.
Here $X_{i}$ and $Z_{i}$ are spin operators in $x$ and $z$ direction, respectively and $N=2 m $ is the number of particles. 
The Hamiltonian can be rewritten as 
\begin{align}
\label{H_Ising_state}
\textstyle{
H=-4J \sum_{i} \varrho_{i,i+1}
+4J \sum_{i} \varrho_{X_{i}}
- 2B \sum_{i} \varrho_{Z_{i}},
}
\end{align}
up to a term proportional to identity, where $i$ runs from $-m$ to $m$.  Here $\varrho_{i,i+1}=\varrho_{X_{i}}\otimes \varrho_{X_{i+1}}$ is a two-qubit state, $\varrho_X=|+\rangle\langle +|=(\mathbbmss{I}+X)/2$ and $\varrho_Z= |0\rangle\langle 0|=(\mathbbmss{I}+Z)/2$, with $|+\rangle= (|0\rangle+|1\rangle)/\sqrt{2}$.
To simulate the imaginary-time evolution of this Hamiltonian, according to the decomposition \eqref{Trotter-Suzuki}, we need to simulate three different types of terms, i.e.,  
$
 U_{i,i+1}=e^{\delta_1\varrho_{i, i+1}} \otimes \mathbbmss{I}_{\bar{i},\widebar{i+1}}
 $, 
 $U_{X_{i}}=e^{\delta_2  \varrho_{X_{i}}} \otimes \mathbbmss{I}_{\bar{i}}$,  and
 $U_{Z_{i}}=e^{\delta_3  \varrho_{Z_{i}}} \otimes \mathbbmss{I}_{\bar{i}}$,
with $\delta_1=\delta_2= 4J \beta/N $ and $\delta_3= 2B \beta/N$, 
such that $e^{-\frac{\beta}{N} H} \approx \prod_{i=-m}^{m} U_{Z_{i}} U_{X_{i}} U_{i,i+1}$. To make sure that initial state of the simulator has a nonzero overlap with the ground state of the Hamiltonian, we assume it is in a uniform superposition of all basis states in the Hilbert space,
$
|\sigma_0\rangle= \otimes_{i=-m}^{m}~|+\rangle_{i}
$.
 Simulation of $U_{Z_{i}}$ and $U_{X_{i}}$ is straightforward. In simulation of $U_{i,i+1}$ the only subtlty is in application of the controlled-\textsc{swap} gate. In this case, because $\varrho_{i,i+1}$ is a two-qubit state, when the control qubit is $|1\rangle$ the swap, which is applied on $\varrho_{i,i+1}$ and the simulator, is multiplication of two ordinary swaps. Equivalently, we can apply two ordinary controlled-\textsc{swap} gates consecutively on a single control qubit and different target systems (first, on $\varrho_{X_i}$ and simulator, second, on $\varrho_{X_{i+1}}$ and simulator).
In Fig.~\ref{fig:fidelity} the fidelity of the simulated state with the exact ground state has been compared with the fidelity of the exact imaginary-time dynamics and the ground state when $i=2$ (the total number of particles is $4$) and $J=1$ for two different cases: close to the degenerate point with $B=0.1$ and far from the degenerate point at $B=5$. 

\textit{Conclusion.---}We have provided a quantum algorithm for ground-state simulation of a Hamiltonian based on imaginary-time evolution. To simulate this nonunitary evolution we have used the SBQS technique.  SBQS has already been shown to be powerful in simulation of state-history dependent Hamiltonians \cite{ECQT-Tavanfar, ECQT}.
Here we have used it to simulate the imaginary-time evolution which is generated by a given Hamiltonian. For this, we first have decomposed the Hamiltonian in terms of a set of resource states that can be prepared in the lab. This way, the focus of simulation changes from interaction generation to state preparation. By using the Trotter-Suzuki expansion the total evolution is decomposed to some steps each of which is generated by a single resource state. By preparing the corresponding resource states and applying controlled-\textsc{swap} gates and measurements appropriately, we have shown how to generate the desired imaginary-time evolution within a given accuracy.  

The runtime (number of steps) of our algorithm is inversely proportional to simulation error and depends quadratically on the number of necessary resource systems. 
Interestingly, size of the system does not explicitly show up in the runtime;
so does the locality of the Hamiltonian. This is because the SBQS technique hinges upon state preparation rather than interactions. 
We have also shown that by deferring the necessary measurements for simulation of each step to the end of the process, we can exponentially improve the success probability as a function of the number of the resource states.

Our method can have applications in qubit initialization, purification, error correction, and novel algorithmic cooling methods. 

\textit{Acknowledgment}.---S. A. would like to thank A. T. Rezakhani for useful comments. 

%%%%%%%%%%%%%%%%%%%%%%%%%%%%%%%%%%%%%%%%%%%%%%%%%%%%%%%%%%%%%%%%

%%%%%%%%%%%%%%%%%%%%%%%%%%%%%%%%%%%%%%%%%%%%%%%%%%%%%%%%%%%%%%%%

\onecolumngrid

%%%%%%%%%%%%%%%%%%%%%%%%%%%%%%%%%%%%%%%%%%%%%%%%%%%%%%%%%%%%%%%%
\clearpage
\appendix

\section{Supplemental Material}
\section{Error of multiplication of erroneous operators}
\label{appendix-error1}
\textbf{Unitary case:} If $U_{1}$, $U_{2}$, $U'_{1}$, and $U'_{2}$ are unitary operators and $\Vert\cdot\Vert$ is the operator norm (induced by the vector norm), then
\begin{equation}
\Vert U_{2}U_{1} -U'_{2}U'_{1}\Vert\leqslant \Vert U_{2}-U'_{2}\Vert + \Vert U_{1}-U'_{1}\Vert.
\label{eq:NC-chaining}
\end{equation}
\textit{Proof.} By using the submultiplicativity property we have
\begin{eqnarray*}
\Vert U_{2}U_{1}-U'_{2}U'_{1}\Vert = \Vert (U_{2}U_{1}-U'_{2}U_{1})+(U'_{2}U_{1}-U'_{2}U'_{1})\Vert &\leqslant& \Vert (U_{2}-U'_{2})U_{1}\Vert + \Vert U'_{2}(U_{1}-U'_{1})\Vert\\
& =&\Vert U_{2}-U'_{2}\Vert + \Vert U_{1}-U'_{1}\Vert,
\end{eqnarray*}
where in the last step we used the unitary invariance of the operator norm.

\textbf{Corollary:} $\Vert \Pi_{k=1}^K U_k -\Pi_{k=1}^K U'_k\Vert \leqslant \sum_{k=1}^K \Vert U_k -U'_k\Vert\leqslant K\cdot\max_k \Vert U_k -U'_k\Vert$ (special case: $\Vert {U}^K-{U'}^K\Vert\leqslant K\Vert U-U'\Vert$). I.e., the error grows at most linearly in the number of unitaries \cite{book:Nielsen-Chuang}.

—Note that
\begin{align*}
U_{3}U_{2}U_{1}-U'_{3} U'_{2}U'_{1} & = U_{3}U_{2}U_{1} - U'_{3}U_{2}U_{1} + U'_{3}U_{2}U_{1} - U'_{3}U'_{2}U'_{1}\\
& =(U_{3} - U'_{3})U_{2}U_{1} + U'_{3} (U_{2}U_{1} - U'_{2}U'_{1})
\end{align*}
and
\begin{align*}
U_{4}U_{3}U_{2}U_{1}-U'_{4}U'_{3} U'_{2}U'_{1} & = U_{4}U_{3}U_{2}U_{1} - U'_{4}U_{3}U_{2}U_{1} + U'_{4} U_{3}U_{2}U_{1} - U'_{4} U'_{3}U'_{2}U'_{1}\\
& =(U_{4} - U'_{4})U_{3} U_{2}U_{1} + U'_{4} (U_{3} U_{2}U_{1} - U'_{3} U'_{2}U'_{1}).
\end{align*}

\textbf{Nonunitary case:} Now assume that $U$s and $U'$s are not necessarily unitary gates but some quantum operations. In that case we have
\begin{align*}
\Vert U_{2}U_{1}-U'_{2}U'_{1}\Vert &\, \leqslant \Vert (U_{2}-U'_{2})U_{1}\Vert + \Vert U'_{2}(U_{1}-U'_{1})\Vert\\
&\, \leqslant \Vert U_{1}\Vert \,\epsilon_{2} + \Vert U'_{2}\Vert \,\epsilon_{1}, \\
& \, \leqslant (\Vert U_{1}\Vert + \Vert U'_{2}\Vert ) \,\epsilon, 
\end{align*}
where $\Vert U'_{k} - U_{k}\Vert \leqslant \epsilon_{k}$ and $\epsilon=\max_{k\in\{1,\ldots,K\}}\epsilon_{k}$. If we also set $M=\max_{k}\{\Vert U_{k}\Vert, \Vert U'_{k}\Vert\}_{k=1}^{K}$, then
\begin{align*}
\Vert U_{2}U_{1}-U'_{2}U'_{1}\Vert &\leqslant  2M \,\epsilon, \\
\Vert U_{3}U_{2}U_{1}-U'_{3}U'_{2}U'_{1}\Vert &\leqslant 3M^{2}\,\epsilon,\\
\Vert U_{4} U_{3}U_{2}U_{1} - U'_{4}U'_{3}U'_{2}U'_{1}\Vert &\leqslant 4M^{3}\,\epsilon.
\end{align*}
By induction,
\begin{align}
\big\Vert \Pi_{k=1}^{K} U_{k} -\Pi_{k=1}^{K} U'_{k} \big\Vert \leqslant KM^{K-1}\,\epsilon.
\label{eq:UU-U'U'-error}
\end{align}
Unlike the case of unitary operators, where $M=1$ and the error bound scales linearly with the number of gates $K$, this error bound is exponential in the number of operations $K$. Note, however, that this exponential bound in some special cases might be too conservative.

\section{Estimation of the error of $\Pi_i e^{-\delta_i A_i} \approx \mathbbmss{I}-\sum_i \delta_i A_i $}
\label{appendix-error2}

In our simulation process, we first approximate $e^{-\delta_i A_i} \approx \mathbbmss{I}-\delta_i A_i$, which allows us to simulate it using a control-qubit in a superposition state and application of a control swap gate. To simulate a multiplication of such processes $\Pi_i e^{-\delta_i A_i}$ we can approximate it with $\Pi_i e^{-\delta_i A_i} \approx \Pi_{i=1}^n (\mathbbmss{I}-\delta_i A_i)$ whose error can be obtained using the method in the last section and is of the order of $O(n \max_i \|e^{-\delta_i A_i}\|^{n-1} \delta_i^2)$. 
Such an approximation means that for simulation we need to repeat the simulation task for each term as an independent unit. However, one may think that if multiplication can be approximated as $\Pi_i e^{-\delta_i A_i} \approx \mathbbmss{I}-\sum_i \delta_i A_i $, it becomes possible to prepare a global multi qubit state and simulate the process by applying controled-$\textsc{swap}$ gates and a global measurement at the end. To estimate the error of this approximation we first estimate the error of approximation $\textstyle{\Pi_{i=1}^n} (\mathbbmss{I}-\delta_i A_i)\approx \mathbbmss{I}-\textstyle{\sum_{i=1}^n} \delta_i A_i $. To do so, we note that 

\begin{align}
(\mathbbmss{I}-\delta_1 A_1)(\mathbbmss{I}-\delta_2 A_2)=&\, \mathbbmss{I}-\delta_1 A_1 -\delta_2 A_2 + \delta_1 \delta_2  A_1A_2 \nonumber\\
(\mathbbmss{I}-\delta_1 A_1)(\mathbbmss{I}-\delta_2 A_2)(\mathbbmss{I}-\delta_3 A_3)=&\, \mathbbmss{I}-\delta_1 A_1 -\delta_2 A_2- \delta_3 A_3 +  \delta_1 \delta_2  A_1A_2+\delta_1 \delta_3  A_1A_3+\delta_2 \delta_3  A_2 A_3 - \delta_1 \delta_2 \delta_2  A_1A_2 A_3\nonumber\\
&\vdots\nonumber
\end{align}
By induction it can be concluded that 
\begin{align}
\textstyle{\Pi_{i=1}^n} (\mathbbmss{I}-\delta_i A_i)=&\, \textstyle{\mathbbmss{I}- \sum_i \delta_i A_i + \sum_{i < j}^n \delta_i \delta_{j} A_i A_{j} - \sum_{i < j<k}^n \delta_i \delta_{j} \delta_k A_i A_j A_k }+ \cdots
\end{align}
From which we obtain
\begin{align}
\|\textstyle{\Pi_{i=1}^n} (\mathbbmss{I}-\delta_i A_i) - \mathbbmss{I} + \textstyle{\sum_{i=1}^n} \delta_i A_i \| \leqslant O( \textstyle{\binom{n}{2}}|\delta|^2 \max_i \|A_i\|^2),
\end{align}
and thus 
\begin{align}
\textstyle{\Pi_{i=1}^n} (\mathbbmss{I}-\delta_i A_i) = \mathbbmss{I}-\textstyle{\sum_{i=1}^n} \delta_i A_i + O( \textstyle{\binom{n}{2}}|\delta|^2 \max_i \|A_i\|^2).
\end{align}
Hence the error of approximation $\Pi_i e^{-\delta_i A_i} \approx \mathbbmss{I}-\sum_i \delta_i A_i $ is estimated as 
\begin{align}
O(n \max_i \|e^{-\delta_i A_i}\|^{n-1} |\delta|^2) +O( \textstyle{\binom{n}{2}}|\delta|^2 \max_i \|A_i\|^2).
\end{align}
%%%%%%%%%%%%%%%%%%%%%%%%%%%%%%%%%5
\section{Proof that we can always assume $H$ is positive}
If $H$ is not positive, it means that its ground state energy $E_0$ is also negative. 
Since $E_0^2 \leqslant \sum_i E_i^2$ therefore $|E_0| \leqslant \sqrt{\sum_i E_i^2}$ or equivalently  $|E_0| \leqslant \|H\|_2$. Hence, $0= E_0+| E_0| \leqslant E_0+  \|H\|_2$. From this, it is easy to conclude that $H+ \|H\|_2 \mathbbmss{I}$ is a positive operator whose ground state is the same as the ground state of $H$. For a positive $H$ there is for sure a decomposition in terms of states with positive coefficients. 

\section{Bounds on Bures distance and fidelity} 
Bures distance $D$ is defined as $D^2=(1-\sqrt{F})$ (which differs modulo an unimportant prefactor $\sqrt{2}$ with the standard definition) and the initial fidelity is given by $F_0= \langle E_0|\sigma_0|E_0\rangle$. According to these definitions, it is easy to see that the Bures distance between the simulated state and the ground state is upper bounded by
\begin{align}
\label{distance}
D(\widetilde{\sigma}_{N\ell}, |E_0\rangle\langle E_0|) &\leqslant D(\widetilde{\sigma}_{N\ell},\widetilde{\sigma}(\beta))+D(\widetilde{\sigma}(\beta), |E_0\rangle\langle E_0|)\nonumber\\
&=D\big(\widetilde{\sigma}_{N\ell} , \widetilde{\sigma}(\beta)\big) +\sqrt{1-\sqrt{F(\widetilde{\sigma}(\beta), |E_0\rangle\langle E_0|)}}= D\big(\widetilde{\sigma}_{N\ell} , \widetilde{\sigma}(\beta)\big) + \sqrt{1- e^{-2\beta E_0} F_0/\mathcal{N}}.
\end{align}
When $\sigma_0$ is the very ground state $F_0=1$  and $\mathcal{N}=e^{-2\beta E_0}$. Since for any $E_i$ we have $ e^{-2\beta (E_i-E_0)} \leqslant  e^{-2\beta \Delta}$ it can be concluded that $\sum_{i\neq 0} e^{-2\beta (E_i-E_0)} \langle E_i|\sigma_0|E_i\rangle \leqslant 
 e^{-2\beta \Delta} \sum_{i\neq 0} \langle E_i|\sigma_0|E_i\rangle$ and $\sum_{i\neq 0} \langle E_i|\sigma_0|E_i\rangle =  \sum_{i} \langle E_i|\sigma_0|E_i\rangle -  \langle E_0|\sigma_0|E_0\rangle = \mathrm{Tr}[\sigma_0] -F_0 =1- F_0$ and hence
\begin{align}
e^{-2\beta E_0} F_0/\mathcal{N} &=\textstyle{1/ \left(1+ \sum_{i\neq 0} e^{-2\beta (E_i-E_0)} \langle E_i|\sigma_0|E_i\rangle/F_0 \right)} \nonumber\\
&
\geqslant \left(1+ e^{-2\beta \Delta} \frac{1-F_0}{F_0} \right)^{-1}.
\end{align}
Thus
\begin{align}
\label{supp13}
\sqrt{1-e^{-2\beta E_0} F_0/\mathcal{N}} & \leqslant \sqrt{ 1-\left(1+ e^{-2\beta \Delta} \frac{1-F_0}{F_0} \right)^{-1}}.
\end{align} 
From this and Eq.~\eqref{distance} we obtain 
\begin{align}
\label{DD}
D(\widetilde{\sigma}_{N\ell}, |E_0\rangle\langle E_0|) &\leqslant D(\widetilde{\sigma}_{N\ell},\widetilde{\sigma}(\beta)) + \sqrt{ 1-\left(1+ e^{-2\beta \Delta} \frac{1-F_0}{F_0} \right)^{-1}}.
\end{align}
Replacing $D\big(\widetilde{\sigma}_{N\ell} , \widetilde{\sigma}(\beta)\big)$ from Eq.~(15) in the main text we obtain
\begin{align}
\label{err}
D(\widetilde{\sigma}_{N\ell}, |E_0\rangle\langle E_0|) &\leqslant \frac{2}{N} \ell^2 \beta^2 h^2 +  \sqrt{ 1-\left(1+ e^{-2\beta \Delta} \frac{1-F_0}{F_0} \right)^{-1}}.
\end{align}
The above relation shows that to achieve an estimate of the ground state to a desired error $\epsilon$, it is sufficient to have
\begin{align}
\frac{2}{N} \ell^2 \beta^2 h^2 +  \sqrt{ 1-\left(1+ e^{-2\beta \Delta} \frac{1-F_0}{F_0} \right)^{-1}}\leqslant \epsilon.
\end{align}

If we want to obtain a bound on fidelity, using the relation $F=(1-D^2)^2$ and the above bound in Eq.~\eqref{DD} we obtain
\begin{align}
F(\widetilde{\sigma}_{N\ell}, |E_0\rangle\langle E_0|) &\geqslant \left[1-\left(\epsilon +\sqrt{ 1-\left(1+e^{-2\beta \Delta} (1-F_0)/ F_0 \right)^{-1}}\right)^2 \right]^2.
\end{align}

%%%%%%%%%%%%%%%%%%%%%%%%%%%%%%%%%%%%%%%%%%%%%%%%%%%%%
\section{Ground state of Ising model in transverse field}

The Hamiltonian of Eq.~\eqref{H_Ising}  is an exactly solvable model whose ground state can be obtained using Jordan-Wigner transformation and is given by 
\begin{align}
|E_0 \rangle=\otimes_{{\ell}=1}^{m} \left[\cos \theta_{\ell} |0\rangle_{-\ell} |0\rangle_{\ell}+i \sin \theta_{\ell}|1\rangle_{-\ell} |1\rangle_{\ell}\right]
\end{align}
where $ |0\rangle_{\pm\ell}$ and  $|1\rangle_{\pm\ell}$ are Bogoliubov fermionic states and 
\begin{align}
\sin 2\theta_{\ell}=\frac{
J \sin \left(\frac{2 \pi \ell}{2m+1}\right)
}
{
\sqrt{
\left(J \cos\left(\frac{2 \pi \ell}{2m+1}\right)-B\right)^2 + J^2 \sin^2 \left(\frac{2 \pi \ell}{2m+1}\right)
}
}.
\end{align}
\end{document}